\documentclass[11pt]{article}
\usepackage{amsmath,amsfonts, amssymb, braket, bbold}
\usepackage{tikz}
\usetikzlibrary{trees,er,snakes,shapes,mindmap}
\usepackage{titlesec}
\titleformat{\section}
 {\normalfont\fontfamily{put}\fontsize{12pt}{16pt}\bfseries\color{black}}
{\thesection}{1em}{}
\def \beq  {\begin{equation}}
\def \eeq  {\end{equation}}
\def \beqar {\begin{eqnarray}}
\def \eeqar {\end{eqnarray}}
\allowdisplaybreaks
\def\sqr#1#2{{\vcenter{\vbox{\hrule height.#2pt
\hbox{\vrule width.#2pt height#1pt \kern#1pt
\vrule width.#2pt}\hrule height.#2pt}}}}

\def\vx {{\vec x}}

\def\vf {{\varphi}}

\def\Tr {{\rm Tr}}

\def\bw {\bar{w}}

\def\vx {{\vec x}}

\def\del {\partial}
\def\bdel{\bar{\partial}}

\def\bz {{\bar{z}}}

\def\A {{\cal A}}

\def\C {{\cal C}}

\def\E {{\cal E}}
\def\F {{\cal F}}

\def\M{{\cal M}}

\def\T {{\cal T}}

\def\vf {{\varphi}}

\def\half{\textstyle{1\over 2}}

\mathchardef\mhyphen="2D
\begin{document}
\fontfamily{put}\fontsize{12pt}{17pt}\selectfont
\def \CMP {{Commun. Math. Phys.}}
\def \PRL {{Phys. Rev. Lett.}}
\def \PL {{Phys. Lett.}}
\def \NPBProc {{Nucl. Phys. B (Proc. Suppl.)}}
\def \NP {{Nucl. Phys.}}
\def \RMP {{Rev. Mod. Phys.}}
\def \JGP {{J. Geom. Phys.}}
\def \CQG {{Class. Quant. Grav.}}
\def \MPL {{Mod. Phys. Lett.}}
\def \IJMP {{ Int. J. Mod. Phys.}}
\def \JHEP {{JHEP}}
\def \PR {{Phys. Rev.}}
\def \JMP {{J. Math. Phys.}}
\def \GRG{{Gen. Rel. Grav.}}
\begin{titlepage}
\null\vspace{-62pt} \pagestyle{empty}
\begin{center}
\vspace{1.3truein} {\Large\bfseries
Topological Terms and Diffeomorphism Anomalies in}\\
\vskip .15in
{\Large\bfseries Fluid Dynamics and Sigma Models}\\
\vskip .5in
{\Large\bfseries ~}\\
{\large\sc V.P. Nair}\\
\vskip .2in
{\sl Physics Department,
City College of the CUNY\\
New York, NY 10031}\\
 \vskip .1in
\begin{tabular}{r l}
{\sl E-mail}:&\!\!\!{\fontfamily{cmtt}\fontsize{11pt}{15pt}\selectfont vpnair@ccny.cuny.edu}\\
\end{tabular}
\vskip 1.5in
\centerline{\large\bf Abstract}
\end{center}
The requirement of diffeomorphism symmetry for the target space can lead to anomalous commutators for the energy-momentum tensor for sigma models and for fluid dynamics,
if certain topological terms are added to the action.
We analyze several examples . A particular topological term is shown to lead to
the known effective hydrodynamics of a dense collection of vortices, i.e. the vortex fluid theory
in 2+1 dimensions. The possibility of a similar vortex fluid in 3+1 dimensions, as well as a fluid
of knots and links, with possible extended diffeomorphism algebras is also discussed.

\end{titlepage}
\fontfamily{put}\fontsize{12pt}{17pt}\selectfont
\pagestyle{plain} \setcounter{page}{2}
\section{Introduction}
The components of the energy-momentum tensor in fluid dynamics or in a field theory will obey
commutation rules which express the fact that they are the generators of diffeomorphisms.
Anomalies in diffeomorphism symmetries  will be reflected, in a Hamiltonian formulation of the
theory, as anomalous commutators. Although we generally seek to avoid such 
anomalies for reasons of unitarity, the following  more nuanced situation can arise.
The fields or fluid variables we are considering are maps from spacetime, denoted as
$M$, into a target manifold $\M$. As is well-known in the context of sigma models,
the choice of local coordinates on $\M$ should not affect physical results, such as the
$S$-matrix. In other words, field redefinitions via diffeomorphisms of $\M$ are possible.
It is then possible that there are certain types of topological terms which can be
included in the action and which can create an incompatibility between diffeomorphisms
in spacetime $M$ and on the target space $\M$. This feature can then be manifest as anomalous commutation rules for the energy-momentum tensor.
Such topological terms are the subject of this paper.

The immediate motivation comes from the work of Wiegmann and Wiegmann and Abanov, who considered vortices in a superfluid, and for the quantum Hall system,
in 2+1 dimensions \cite{WA}. In a situation with a large number of
vortices, it is possible to consider an effective hydrodynamics for them.
In other words, each vortex can be viewed as a point-particle and a fluid with such constituents is
obtained. This fluid is different from the underlying fluid which produced the vortices in the
first place. The authors of \cite{WA} showed that the commutation rules for the energy-momentum tensor for the vortex fluid has anomalous terms.
We may recall that anomalous commutators can be viewed as 2-cocycle terms obtained via the
descent equations from an index density in two higher dimensions, and hence, they are closely tied
to the existence of gravitational anomalies \cite{cocycle}.
Since there are no purely gravitational anomalies in 2+1 dimensions \cite{anom}, how is it possible
to have anomalous commutators? Could they arise from the incompatibility mentioned above?

There is also a larger context for our analysis in view of the recent resurgence of interest
in fluid dynamics. The behavior of a quantum Hall droplet as an incompressible fluid, with the possibility of nondissipative viscosity in 2+1 dimensions \cite{QHE-visc} and the holographic fluid-gravity correspondence in the
AdS/CFT framework \cite{holo} have been two major tracks for ongoing research. Added to this is the fact that
a formalism for nonabelian fluid dynamics incorporating anomalous symmetries
\cite{review}
is clearly the natural framework for interesting physical phenomena
 such as the chiral magnetic effect and its variants \cite{ChiMag}.
  And, of course, the fluid version of the
 Wess-Zumino term as an effective action for anomalies
 is the classic example of a topological term which can influence the dynamics of a fluid \cite{{NRR},{AMN}}.
 In the present analysis, we will focus on a 
slightly different class of topological terms.
 We will consider terms which can lead to anomalous commutators
as well as  terms which can couple different fluids.
 Notice that the example of the vortex fluid may be considered as a two-fluid system
 with the fundamental underlying fluid and the vortex fluid, 
 so it should be interesting to analyze systems with
 independent dynamics for each component except for
 coupling via topological terms.
 
 A useful observation is that anomalous commutators define 2-cocycles in the operator algebra
 \cite{anom}.
 For the equal-time algebra for a theory in 3+1 dimensions, we should thus consider
 a 5-form which is closed but not exact.
 Locally such a 5-form can be written as the exterior derivative
 of a 4-form $\Gamma$. This can be added to the action, and can lead to 
 anomalous commutators. Thus, in 3+1 dimensions,
 our strategy will be to consider sigma models or fluid variables for which
we can identify nontrivial 5-forms.
 For 2+1 dimensions, we will need nontrivial 4-forms.
 
 This paper is organized as follows. In section 2, we analyze the sigma model with target space
 $\mathbb{CP}^2$, showing how the extended version of the diffeomorphism algebra 
 arises and how it is connected to diffeomorphism of the target space.
 In section 3, we consider various types of topological terms which can be added to the standard
 action for fluid dynamics. This is done in terms of a group-theoretic formulation of
 the Clebsch variables, which helps to simplify the analysis.
 Section 4 is devoted to the case of one of the topological terms and the corresponding
 extended version of
 the diffeomorphism algebra is obtained. In section 5, we carry out the necessary comparison to identify this case with the vortex fluid work of \cite{WA} in 2+1 dimensions, and also show that a special case yields a central extension identified in \cite{Raj} for 3+1 dimensions.
 In section 6, we analyze the other topological term, designated $I_2$,
 and argue that the extended algebra obtained may apply for an effective hydrodynamics
 of knots and links in 3+1 dimensions. The paper concludes with a short discussion.

\section{A sigma model on $\mathbb{CP}^2$}
We will start with a sigma model in 2+1 dimensions with the target space
$\M$ as the complex projective space $\mathbb{CP}^2$. This will serve as a concrete example
which sets the paradigm for later discussion.
The space $\mathbb{CP}^2$ has nontrivial $H^4$ and a generating element of this can be taken
as $\Omega\wedge \Omega$, where $\Omega$ is the K\"ahler two-form.
We can think of $\mathbb{CP}^2$ as $SU(3)/U(2)$ and use 
a group element $U \in SU(3)$ with the identification
$U \sim U\,h$, $h \in U(2) \subset SU(3)$ to coordinatize the manifold.
In a $3\times 3$ matrix representation of $U$, the K\"ahler one-form is given by
\beq
A = i {2\over \sqrt{3}} \Tr ( t_8 U^{-1} dU) , \hskip .2in
t_8 = {1 \over 2 \sqrt{3}} \left[ \begin{matrix} 1&0&0\\ 0&1&0\\ 0&0&-2\\ \end{matrix}
\right]
\label{diff1}
\eeq
Under $U \rightarrow U\, h$, $A$ is not invariant, but transforms as
\beq
A (U h) = A (U) - {1\over \sqrt{3}} d \theta_8
\label{diff2}
\eeq
where $h = \exp( i t_8 \theta_8 + i t_i \theta_i )$, $i = 1,,2, 3$.
Thus $A$ is not a one-form on the coset $SU(3)/U(2)$, but K\"ahler two-form
$\Omega = d A$ is invariant under $U \rightarrow U h$ and is well defined
on $\mathbb{CP}^2$. One can introduce local coordinates for
the manifold by writing
\beq
U_{i3} = {1\over \sqrt{1+ \bz \cdot z} } \bigl( z^1, z^2, 1\bigr)
\label{diff3}
\eeq
It is also useful to consider real coordinates defined by, say,
$z^1 = \vf^1 + i \vf^2 $, $z^2 = \vf^3 + i \vf^4$. In terms of these parametrizations, the 
one-form $A$ can be written as
\beqar
A &=& - {i \over 2} \,{ \bz \cdot dz - z\cdot d\bz \over (1+ \bz \cdot z)}
= - J_{ab} {\vf^a d\vf^b \over (1+ \vf^2)}\nonumber\\
J_{12} &=&- J_{21} = 1, \hskip .2in J_{34} = - J_{43} = 1, \hskip .2in {\rm all~other~} J_{ab} = 0
\label{diff4}
\eeqar
The product $\Omega \wedge \Omega$ can be written as $d \Gamma$ where
$\Gamma = A \wedge d A$. While $\Omega \wedge \Omega$ is well defined on
$\mathbb{CP}^2$, $\Gamma$ does not descend to the coset space since 
$A$ is not invariant under $U \rightarrow U h$. $\Gamma$ will be the topological term
we add to the action. Thus the theory we are considering is defined by the action
\beq
S = {1\over 2} \int G_{ab}\, \del_\mu \vf^a \del^\mu \vf^b 
+ k \int \Gamma
\label{diff5}
\eeq
where $k$ is a constant and $G_{ab}$ is the metric tensor for
the target space $\mathbb{CP}^2$. Notice that $\Gamma$ shifts by a
total derivative under $U \rightarrow U h$, so that the bulk action is
well-defined with appropriate boundary conditions. We take the fields to vanish at
spatial infinity.
The surface terms on equal-time spatial slices do not necessarily
vanish and can lead to a canonical transformation.
 In terms of the real coordinates $\vf^a$, $\Gamma$ is given by
\beqar
\Gamma &=& {1\over 3} \epsilon_{abcd} {\vf^a d\vf^b d\vf^c d\vf^d \over (1+ \vf^2 )^2}
\nonumber\\
&=&  {1\over 3} \epsilon_{abcd} {\vf^a \del_\mu\vf^b \del_\nu\vf^c \del_\alpha\vf^d \over (1+ \vf^2 )^2}
\, dx^\mu \wedge dx^\nu \wedge dx^\alpha
\label{diff6}
\eeqar
The canonical momentum can be read off from the action as
\beqar
\Pi_a &=&  G_{ab}\, {\dot \vf}^a  - \Gamma_a \label{diff7}\\
\Gamma_a&=&k \, \epsilon_{abcd} {\vf^b \, d\vf^c \, d\vf^d \over (1+ \vf^2)^2}
= k \, \epsilon_{abcd} {\vf^b \, \del_i\vf^c \, \del_j \vf^d \over (1+ \vf^2)^2}
\, dx^i \wedge dx^j
\label{diff8}
\eeqar
(The differentials $dx$ in $\Gamma_a$ in this equation are for the spatial coordinates only.)

The term $\Gamma$ we have added is a differential form on spacetime 
and is therefore independent of the spacetime metric. Therefore it will not contribute to the
energy-momentum tensor.
By considering the variation of the action with respect to the spacetime
metric, we obtain the
energy-momentum tensor as
\beq
T_{\mu \nu} = G_{ab} \del_\mu\vf^a \del_\nu\vf^b - \eta_{\mu\nu} \, {1\over 2} (G \del\vf \del\vf)
\label{diff9}
\eeq
The momentum density which can be identified as the generator of spatial diffeomorphisms
is given by $T_{i0}= G_{ab} \del_i \vf^a {\dot \vf}^b$.
 The generator of the transformation $x^i \rightarrow x^i + \xi^i$
is thus given by
\beqar
T (\xi ) &=& \int ( \xi^i \del_i \vf^a ) \, G_{ab} \, {\dot \vf}^b
=  \int ( \xi^i \del_i \vf^a ) \,  ( \Pi_a + \Gamma_a )\nonumber\\
&=&  \int ( \xi\cdot \del \vf^a ) \,  \left( - i {\delta \over \delta \vf^a}  +  \Gamma_a \right))
= - i \int ( \xi\cdot \del \vf^a ) \,  D_a\label{diff10}\\
D_a &=& \left( {\delta \over  \delta \vf^a} + i\, \Gamma_a \right)
\nonumber
\eeqar
$D_a$ is a covariant derivative for the target space with $\Gamma_a$ as the gauge field.

It is now completely straightforward to calculate the commutator of two such
generators. We find
\beq
[T (\xi) , T (\xi')] = i T ([\xi,\xi']) - \int \rho^a (x)\, \sigma^b(y)\,
\, [ D_a , D_b] 
\label{diff11}
\eeq
where $[ \xi, \xi']^i = \xi\cdot \del \xi'^i - \xi'\cdot \del \xi^i$
and $\rho^a = (\xi\cdot\del\vf^a)$ and $\sigma^a = (\xi'\cdot \del \vf^b)$.
We can think of $\Gamma_a$ as a connection or gauge field on the space of fields
and hence the commutator $[D_a, D_b]$ is the field strength,
\beq
[D_a, D_b ] = i \, \left( {\del \Gamma_b (y) \over \del \vf^a(x) }
- {\del \Gamma_a (x) \over \del \vf^b(y) }\right)
\equiv i \F_{ab} (x,y) 
\label{diff12}
\eeq
It is simpler to use the notation of differential forms for the target space
and write the connection as
\beq
\A = \int \Gamma_a \delta \vf^a
\label{diff13}
\eeq
where $\delta$ denotes the exterior derivative for the space of fields.
What we need for the curvature (\ref{diff12}) is thus $\delta \A$.
In terms of $\Omega = {\half} \Omega_{ab} d\vf^a d\vf^b$, we can write $\A$ as
\beq
\A = k \int \left[ A_a \delta \vf^a \, \Omega + \Omega_{kl} \delta\vf^k d\vf^l\, A \right]
\label{diff14}
\eeq
(We do not write the wedge sign any more to avoid too much clutter, it is taken as understood.
Notice that the comparison of (\ref{diff13}) and (\ref{diff14}) gives another
expression for $\Gamma_a$ as well.)
To obtain the curvature, we may note the following identities.
\beqar
\delta (A_a \delta \vf^a) &=& \half \Omega_{ab} \delta \vf^a \, \delta \vf^b\nonumber\\
\delta \Omega &=& d ( \Omega_{kl} \delta \vf^k \vf^l )\nonumber\\
\delta ( \Omega_{kl} \delta \vf^k \, d\vf^l ) &=& - d (\half \Omega_{kl} \delta \vf^k \delta \vf^l)
\label{diff15}\\
\delta (A_a d\vf^a )&=& d (A_a \delta \vf^a) + \Omega_{ab} \delta \vf^a d\vf^b
\nonumber
\eeqar
Using these results we can calculate $\delta \A$ as
\beq
\delta \A = k \int \left[ \half \Omega_{ab} \delta \vf^a \, \delta \vf^b \, \Omega_{kl} d\vf^k \, d\vf^l
- \Omega_{ab} \delta\vf^a d\vf^b \, \Omega_{kl} \delta \vf^k d\vf^l
\right]
\label{diff16}
\eeq
Some total derivatives in the integrand
have been dropped since they integrate to zero. We assume the boundary conditions
are such that this is the case. The second term on the right hand side of
(\ref{diff11}) can now be written as
\beqar
- \int \rho^a (x)\, \sigma^b(y)\,
\, [ D_a , D_b]  &=&- i k  \int \left[ \half \Omega_{ab} \rho^a \sigma^b \, \Omega_{kl} d\vf^k \, d\vf^l
- 2 \,\Omega_{ab} \rho^a d\vf^b \, \Omega_{kl} \sigma^k d\vf^l
\right]\nonumber\\
&=& - i k~ V_\rho \rfloor V_\sigma \rfloor \F
\label{diff17}
\eeqar
where the symbol $V_\rho\rfloor$ denotes the interior contraction with the functional vector
field
\beq
V_\rho = \int \rho^a {\delta \over \delta \vf^a}
\label{diff18}
\eeq
Explicitly, for a (functional) differential form $F = \int \half F_{ab} \delta\vf^a \delta \vf^b$,
\beq
V_\rho\rfloor F = \int \rho^a {\delta \over \delta \vf^a} \rfloor \int \half F_{ab} \delta\vf^a \delta \vf^b
= \half\int  \left[ F_{ab} \rho^a \delta \vf^b  - F_{ab} \delta \vf^a \rho^b\right]
= \int  F_{ab} \rho^a \delta \vf^b
\label{diff19}
\eeq
Consider now the differential 4-form $\Omega^2$ on the target space. 
We do contractions with $V_\rho$ and $V_\sigma$ and write it as a differential
form on space
by taking $\vf^a$ as functions of the coordinates; i.e., we pull back the result
to spatial manifold. We can then easily check that
\beq
V_\rho\rfloor V_\sigma\rfloor \F = V_\rho\rfloor V_\sigma\rfloor (\Omega\, \Omega )
\label{diff20}
\eeq
We can now rewrite (\ref{diff11}) for the
 commutator of the generators of spatial diffeomorphisms as
\beq
[T(\xi) , T(\xi') ] = i \, T([\xi, \xi']) - i k \int V_\rho\rfloor V_\sigma\rfloor (\Omega^2)
\label{diff21}
\eeq
The first term on the right hand side is what is expected from the fact that a diffeomorphism
$x^i \rightarrow x^i + \xi^i$
on a space-dependent function $f$ leads to the change
\beq
\delta f = \xi^i {\del \over \del x^i} \, f(x)
\label{diff22}
\eeq
Th second term on the right hand side of (\ref{diff21}) shows that the generators $T(\xi )$
for the diffeomorphisms have anomalous commutation rules, the anomaly being related
to the $H^4$ element of the target space $\M$. 
The definition of the generator as in (\ref{diff10}) also shows how this anomaly can be avoided.
Define
\beq
\T (\xi) =  T(\xi ) - \int (\xi\cdot \del \vf^a) \, \Gamma_a
= -i \int (\xi\cdot \del \vf^a ) {\delta \over \delta \vf^a}
\label{diff23}
\eeq
It is then trivial to see that $[\T (\xi), \T (\xi') ] = i \T ([\xi, \xi'])$, with no anomalous terms.
However, $\T$ is related to the components of the energy-momentum tensor
via the subtraction of the integral of $ (\xi\cdot \del \vf^a) \, \Gamma_a$. Since 
$\Gamma_a$ is not well defined on $\mathbb{CP}^2$, as we have mentioned after
(\ref{diff2}), this redefinition is problematic. In other words, $\A$ is a gauge field
on the space of field configurations and hence not invariant under field redefinitions or
target space diffeomorphisms. Thus, while the use of
$\T (\xi)$ will eliminate the anomaly for diffeomorphisms of the spatial manifold,
we lose the freedom of field redefinitions or target space diffeomorphisms.

This is the key result of this section. 
We can add to the action a term $\int \A_a {\dot \vf}^a$ where
$\A$ is the potential for an element of $H^4$ (or $H^{d+1}$ for $d$-dimensional
spacetime) of the target space. This can lead to a conflict between diffeomorphisms of the
base spatial manifold and the space of field configurations, resulting in anomalous
commutators.
In  the next two sections, we will explore a similar structure for fluids
in 2+1 and 3+1 dimensions.
\section{The nature of possible topological terms for fluids}
We start with fluids in 3+1 dimensions; the 2+1 dimensional case can be easily obtained
by a simple reduction.

In the classic Lagrange approach to fluid dynamics, one considers a multiparticle system,
where $x^i (z, t)$ denotes the position of the $z$-th particle at time $t$, where
$z$ is an element of some indexing set labeling the particles.
When the number of particles is very large and a continuum approximation 
is meaningful, one chooses the initial positions of the particles as the label for the
particle. In other words, $x^i (z, 0) = z^i$. Thus $x^i (z, t)$ may be regarded as the image
of $z^i$ under a diffeomorphism parametrized by the time coordinate $t$.
The kinetic term in the action takes the usual form 
\beq
S_{\rm kin}  = {1\over 2} \int d^3z\, \rho_0 (z) \, {\dot x}_i {\dot x}^i
\label{diff24}
\eeq
We take the particle mass to be $1$ and $\rho_0 (z)$ gives the number density of particles
as a function of the fiducial variables $z^i$.
The canonical one-form at the level of particles is obviously given by
\beq
\A = \int d^3z \rho_0(z)\, v_i \delta x^i
= \int d^3z \rho_0(z)\, {\dot x}_i \delta x^i
\label{diff25}
\eeq
The use of the notation  $\delta x^i$ rather than $dx^i$ signifies that this is to be viewed
as a one-form
on the space of configurations.
If the helicity of the fluid system is fixed, then the velocity admits
the Clebsch parametrization
\beq
v_{i}=\partial_{i}\theta+\alpha\, \partial_{i}\beta
\label{diff26}
\eeq
for an arbitrary functions $\theta,\alpha,\beta$. The canonical one-form, for this parametrization, reduces to
\beqar
\A &=& \int d^3z \rho_0 (z)\, (\del_i \theta + \alpha \del_i \beta ) \, \delta x^i
= \int d^3z \rho_0 (z)\, (\delta\theta + \alpha \delta \beta )\nonumber\\
&=& \int d^3x \rho (x)\, (\delta\theta + \alpha \delta \beta )
\label{diff27}
\eeqar
where the density $\rho (x)$, as a function of the $x$-coordinates, is defined by
$ d^3z \rho_0 (z) = d^3x \rho (x)$.
This shows that
to obtain the canonical one-form as in (\ref{diff27})
we should take 
 the term in the action involving time-derivatives
to be $\int \rho\,\dot{\theta}+\rho\,\alpha\,\dot{\beta}$.
A suitable action for fluid dynamics (in terms of the Eulerian variables)
 is then
\begin{equation}
S=\int\rho\,\dot{\theta}+\rho\,\alpha\,\dot{\beta}-\left[\frac{1}{2}\,\rho\,{v}^{2}-V\right].
\label{diff28}
\end{equation}
Here we have also included a term corresponding to the potential energy.
This expression gives the action suitable for the
Clebsch parametrization with
$(\rho,\theta)$, $(\rho\alpha,\beta)$ forming two sets
of canonically conjugate variables.

There is a group-theoretic version of the Clebsch parametrization which is
also useful. Towards this, consider the group $SU(1,1)$.
A typical element $g$ may be parametrized as
\beq
g =\frac{1}{\sqrt{1-\bar{u}u}}\left[\begin{array}{cc}
1 & u\\
\bar{u} & 1
\end{array}\right]\left[\begin{array}{cc}
e^{i\theta/2} & 0\\
0 & e^{-i\theta/2}
\end{array}\right]
\label{diff29}
\eeq
where $u$ is a complex variable. It is easy to verify that
\beqar
-i \Tr \left(\sigma_{3}\,g^{-1}\,dg\right) &=& d\theta+\alpha \,d\beta\nonumber\\
\alpha &=& \frac{\bar{u}u}{1-\bar{u}u}, \hskip .2in \beta=\left(-i/2\right)\ln\left(\frac{u}{\bar{u}}\right).
\label{diff30}
\eeqar
The variable $\theta$ corresponds to the compact direction,
or $U(1)$ subgroup generated by the Pauli matrix $\sigma_3$;
$\alpha$ and $\beta$ parametrize $SU(1,1)/U(1)$. 
The action (\ref{diff28})
can be written in terms of $g$ as
\beq
S=-i\int J^{\mu}\, \Tr\left(\sigma_{3}g^{-1}\partial_{\mu}g\right)-\int\left[\frac{J_{i}J_{i}}{2\rho}+V\right]\label{diff31}
\eeq
where we denote $J^{0}=\rho$. 
$J_i$ can be eliminated by its equation of motion and leads back to the form
in (\ref{diff28}).
It is also easy to make a relativistic generalization, with the action given by
\beq
S=-i\int J^{\mu}\, \Tr\left(\sigma_{3}g^{-1}\partial_{\mu}g\right)-F(n)
\label{diff32}
\eeq
where $F(n)$ is a function of the variable $n$, which is defined
by $J^{\mu} J_{\mu}=n^{2}$.
The function $F(n)$ will characterize the fluid\footnote{For a general discussion about using group-theoretic variables for fluid dynamics, see \cite{{review},{NRR}}.}.
We will not discuss this in any more detail, except to note that the 
$T_{i0}$ component of the energy-momentum tensor
for (\ref{diff32}) is given by
\beq
T_{i0} = \rho \, (\del_i \theta + \alpha \del_i \beta )
\label{diff33}
\eeq
Given that $(\rho,\theta)$, $(\rho\alpha,\beta)$ are canonical pairs, we verify easily that
\beq
[ T(\xi ), T(\xi')]= i T([\xi, \xi']), \hskip .2in T(\xi) = \int \xi^i T_{i0}
\label{diff34}
\eeq

Our aim is to consider topological terms which one can add to the action
(\ref{diff31}), or (\ref{diff32}), and which can potentially lead to anomalous commutation rules for
diffeomorphisms.
However, a comment is in order, before we move on.
The compact $U(1)$ direction of the $SU(1,1)$ may be a bit
puzzling, since the
classical Clebsch parametrization does not have a compactness requirement.
Using (\ref{diff27}), we get
\beq
\left[\rho(f),g(x)\right]=-i \,g(x)\,\frac{\sigma_{3}}{2}f(x), \hskip .3in
\rho (f) = \int f(x) \rho(x)
\label{diff35}
\eeq
This means that in the quantum theory
\beq
{\mathcal U}^{\dagger}\,g \,{\mathcal U} = g \, e^{i\pi\sigma_{3}}= - g 
\label{diff36}
\eeq
with ${\mathcal U} =\exp\left[-2\pi i\int\rho\right]$. 
All observables involve
even powers of $g$, and so  are invariant under the action of ${\mathcal U}$.
Effectively, we can set  $\,{\mathcal U} =1$, giving $\int\rho=N$ for some integer
$N$. 
This is equivalent to saying
that the fluid is made of particles with $\rho$ being the number
density. Since this is what happens in reality, we regard the
existence of a compact direction
as a good feature, justifying the use of $SU(1,1)$.
 (If the total vorticity
is also quantized we should use $SU(2)$ in place of $SU(1,1)$.)

Turning to possible topological terms, we consider differential forms we can construct using $g$.
Given $-i \Tr (\sigma_3 g^{-1} dg )$, we can construct the 2-form
\beq
\omega = d (-i \Tr (\sigma_3 g^{-1} dg )) = i \Tr (\sigma_3 (g^{-1} dg )^2)
\label{diff37}
\eeq
The spatial components of this correspond to the vorticity with the identification (\ref{diff30}).
Further, we  have the 3-form
$-i \Tr (\sigma_3 g^{-1} d g) \wedge \omega$, which is proportional to
$\Tr (g^{-1} dg )^3$ for dimensional reasons. The integral of 
$\Tr (g^{-1} dg )^3$ is the helicity of the fluid and is known to commute with all observables
if we use the standard commutation rules for a fluid.
Since $\omega\wedge \omega$ is zero (for dimensional reasons), some of the interesting topological terms we can construct
using $g$ are:
\vskip .05in\noindent
\renewcommand{\arraystretch}{1.2}
\begin{tabular}{l l l}
1. & $ I_1 =\int \omega \wedge B$,\hskip .2in&$B$ = 2-form in 3+1, 1-form in 2+1 dimensions\\
2. & $I_2 = \int \Tr (g^{-1} dg )^3 \wedge C$,&$C$ = 1-form in 3+1, 0-form in 2+1 
dimensions\\
3. &$I_3 = \int \Tr (\sigma_3 g^{-1} dg ) \wedge \Omega $,
&$\Omega$ = 3-form in 3+1, 2-form on 2+1 dimensions\\
\end{tabular}
\renewcommand{\arraystretch}{1}

The first one, namely $I_1$, is easy to dispose of. Since $\omega = d (-i \Tr (\sigma_3 g^{-1} dg ))$, an
integration by parts shows that $I_1$ is a surface term if $B$ is a closed form.
Thus it will not affect the equations of motion or the canonical structure in the bulk.
We will assume boundary conditions such that the surface term is zero.
If $B$ is not a closed form, it reduces to $I_3$, with $\Omega = d B$.

Turning to $I_2$, notice that the variation of $\Tr (g^{-1} dg)^3$ is a total derivative and hence
$I_2$ will not contribute to the equations of motion if $C$ is closed, i.e., $d C = 0$.
By considering the term with time-derivatives of $g$ in $I_2$, we can see that its contribution to the canonical 1-form is
\beq
\Delta \A = - 3 \int \Tr \left[ g^{-1} \delta g \, d (g^{-1} dg ) \right] \wedge C
\label{diff38}
\eeq
This leads to $\delta (\Delta \A ) = -3 \int d \left[ (g^{-1} \delta g )^2 g^{-1} dg \right] \wedge C$,
so that, if $C$ is closed, $I_2$ does not contribute to the canonical 2-form either.
(Again, we assume boundary conditions where the surface term does not contribute.)
In other words $\Delta \A$ is a flat connection on the space of configurations
$\{ g (x) \}$. While it does not affect the Poisson brackets of observables, it does lead to
a vacuum angle (via a term like $\theta \, I_2$), characterizing the state of the fluid in the
quantum theory.

If $C$ is not closed, we can have a nonzero $\delta (\Delta \A )$. In this case, other than an external
field, a natural choice for $C$ would be something like
$\Tr (\sigma_3 h^{-1} d h )$, where $h \in (SU(1,1)$ refers to another fluid.
Thus $I_2$ will be a topological term coupling two fluids.
However it is then of the form $I_3$ with $\Omega = \Tr (h^{-1} d h)^3$ if we further reverse the roles of the two fluids with an
exchange $g \leftrightarrow h$.

The interesting cases to emerge from this analysis are thus: a) $I_3$ with $\Omega$ external \, 
b) $I_2$ with $C$ external, with
$d C \neq 0$ \,  c) $I_2$ with $C \sim \Tr (\sigma_3 h^{-1} dh )$,
which is also the same as $I_3$ with $\Omega = \Tr (g^{-1} dg )^3$ with
an exchange of $g$ and $h$. We will now analyze these cases in some detail below.

\section{The term $I_3$ in 3+1 and 2+1 dimensions}
We now consider a fluid where, in addition to the usual terms, we add
a term proportional to $I_3$ in the action. Thus the action is
taken to be of the form
\beq
S = -i \int \rho\, \Tr (\sigma_3 g^{-1} \del_0 g )  + i k \int \Tr (\sigma_3 g^{-1} d g ) \wedge \Omega
- \int dt \, H
\label{diff39}
\eeq
Here $k$ is a constant and 
$\Omega$ is a 3-form for fluids in 3+1 dimensions, a 2-form in 2+1 dimensions.
If $\Omega$ has a time-component, then $g^{-1} dg$ in the extra term will be a spatial derivative
and will not contribute to the canonical structure. So, for our purpose, we will assume that 
$\Omega$ is a 3-form/2-form on the spatial manifold. Any time-components can be added to our analysis without affecting the canonical structure which is the focus of our work.
For brevity we write
\beq
{\bar \rho} = k \begin{cases}
{1\over 3!} \epsilon^{ijk} \Omega_{ijk} &~~~( 3+1 \, {\rm dimensions})\\
{1\over 2!} \epsilon^{ij} \Omega_{ij} &~~~(2+1\, {\rm dimensions})\\
\end{cases}
\label{diff40}
\eeq
(We are taking the dual of $\Omega$ to get ${\bar\rho}$, so that there should be a factor
of $(\det g_{\mu\nu})^{-1/2}$ in curved space, where $g_{\mu\nu}$ is the metric tensor.)
The action (\ref{diff39}) can be written as
\beq
S = -i \int (\rho- {\bar\rho})\, \Tr (\sigma_3 g^{-1} \del_0 g )  
- \int dt \, H
\label{diff41}
\eeq
The canonical one-form and two-form are given by
\beqar
\A &=& -i \int (\rho- {\bar\rho})\, \Tr (\sigma_3 g^{-1} \delta g )  \nonumber\\
\delta \A &=&- i \int \delta \rho\, \Tr (\sigma_3 g^{-1}\delta g) + i \int (\rho - {\bar\rho})
\Tr (\sigma_3 g^{-1} \delta g\, g^{-1} \delta g )
\label{diff42}
\eeqar
It is now straightforward to work out a number of Poisson brackets. Consider a vector
field $L(\theta )$ which corresponds to left translations on $g$ given by
$V_L (\theta ) \,g = - i \theta^a t_a \, g$. Contracting this vector field with $\delta \A$, we find
\beq
V_L (\theta) \rfloor \delta \A = \delta \left[ \int (\rho - {\bar\rho}) \Tr(\sigma_3 g^{-1} \theta^a t_a g ) \right]
\label{diff43}
\eeq
which corresponds to the Poisson bracket relation
\beq
[ L(\theta ) , g(x)  ] = i \theta^a(x)  t_a \, g(x) , \hskip .3in
L(\theta ) = - \int (\rho - {\bar \rho}) \, \Tr (\sigma_3 g^{-1}  t_a g )\,\theta^a
\label{diff44}
\eeq
In a similar way, it is easy to obtain the relation
\beq
[ \rho (f), g (x) ] = -i g(x)\,t_3\, f(x)
\label{diff45}
\eeq
We now turn to spatial diffeomorphisms given by a vector field $V_\xi$ defined by
\beq
V_\xi \, g = \xi^i \del_i g , \hskip .3in V_\xi \, \rho = \nabla \cdot \bigl[\xi (\rho - {\bar \rho})\bigr]
\label{diff46}
\eeq
(We consider ${\bar\rho}$ to be spatially constant for this.) The contraction of this vector field
with $\delta \A$ gives
\beqar
V_\xi \rfloor \delta \A &=& - \delta \T(\xi )\nonumber\\
\T(\xi)&=& - i \int (\rho -{\bar\rho}) \Tr (\sigma_3 g^{-1} \xi\cdot \del g ) = 
\int (\rho - {\bar \rho}) \, \xi^i v_i
\label{diff47}
\eeqar
where $v_i = -i \Tr (\sigma_3 g^{-1}\del_i g) $ is the fluid velocity as in (\ref{diff26}).
$\T (\xi )$ is thus the canonical generator of diffeomorphisms and
it obeys the Poisson bracket algebra
\beq
[ \T (\xi ), \T (\xi') ] = \T ([\xi,\xi']), \hskip .3in
[\xi, \xi']^i = \xi\cdot \del \xi'^i - \xi'\cdot \del \xi^i
\label{diff48}
\eeq
This algebra is as expected for diffeomorphisms. However, if we define the energy-momentum tensor
by varying the action with respect to the metric, it has no contribution from the topological term, and we find
\beqar
T (\xi ) &=& \int \xi^i T_{i0} = \int \rho\, \xi^i v_i = -i \int \rho\, \Tr (\sigma_3 g^{-1} \xi\cdot \del g)
\nonumber\\
&=& \T (\xi) + \int {\bar \rho}\, \xi^i v_i
\label{diff49}
\eeqar
From the Poisson bracket relations given above, we can easily verify that
\beqar
\int \xi\cdot v \, [ (\rho - {\bar \rho}), \int \xi'^i v_i ]
&=& - \int \xi'\cdot \del (\xi \cdot v)\nonumber\\
{}[v_i (x), v_j (y)] &=& - {1\over \rho - {\bar\rho} } (\del_i v_j - \del_j v_i ) \, \delta (x-y)
\label{diff50}
\eeqar
Using these relations, the Poisson bracket algebra for $T(\xi)$ can be worked out as
\beq
[ T(\xi), T(\xi')] = T ([\xi, \xi']) - \int \left( {\rho {\bar\rho} \over \rho - {\bar\rho}}
\right) \xi^i \, \xi'^j ( \del_i v_j - \del_j v_i )
\label{diff51}
\eeq
We see that the algebra for $T(\xi)$ has an extension term involving the density and the
vorticity $\omega_{ij} = \del_i v_j - \del_j v_i $.
This extension is absent for $\T (\xi )$ which is obtained by adding the integral of
$ - {\bar\rho}\, \xi\cdot v$ to $T(\xi )$.
Since this extra term is well defined on the space of field configurations, the extension
in the algebra (\ref{diff51}) is not a true anomaly. To put this another way,
it is cohomologically trivial, since it can be removed by a redefinition of the generators.
So far this is in keeping with the absence of gravitational anomalies in 3+1 and 2+1 dimensions.

However, we can consider a reduction of the algebra (\ref{diff51}) to the case of an incompressible fluid where we set $\rho - {\bar \rho} $ to some constant $\rho_0$; i.e., we impose a constraint
\beq
\rho - {\bar \rho } - \rho_0 \approx 0
\label{diff52}
\eeq
In the canonical reduction, we need a conjugate constraint, which may be taken as
$\theta \approx 0$, where $\theta$ is the Clebsch variable in
(\ref{diff26}) and (\ref{diff29}). The phase space is reduced to the set of all maps from space
into $SU(1,1)/U(1)$. The fluid velocity $-i \Tr (\sigma_3 g^{_1} \del_i g )$ is not
invariant under a shift of $\theta$ and hence does not descend to
the reduced space.
The addition of the integral of $ - {\bar\rho} \,\xi\cdot v$ to $T(\xi )$ is not defined on the reduced space
and so the extension in (\ref{diff51}) becomes a true anomaly.
This is very similar to what we found for the sigma model in section 2.

Strictly speaking, we should also reduce the Poisson bracket algebra to an algebra for Dirac brackets
to see if there is any change in the extension.
But notice that the Poisson bracket of $T(\xi )$ with the constraint
$(\rho - {\bar \rho } - \rho_0)$ vanishes on the constrained space since
\beqar
[ T(\xi ), \int f (\rho - {\bar \rho } - \rho_0 ) ] &=&
\int \rho\, \xi\cdot \del f = \int (\rho - {\bar \rho } - \rho_0 ) \, \xi\cdot \del f 
- \int f ({\bar \rho} + \rho_0  ) \nabla\cdot \xi\nonumber\\
&\approx& 0
\label{diff53}
\eeqar
for divergence-free vector fields $\xi^i$. (When we consider incompressible fluids
only diffeomorphisms by divergence-free vector fields
are meaningful.)
As a result of this relation, the Dirac bracket $[ T(\xi ), T (\xi') ]_*$ has the same right hand side
as in (\ref{diff51}).
\section{Physical examples of the $I_3$ term}
In this section we will consider specific physically interesting cases for which the
$I_3$ topological term can be used.
\subsection{The vortex fluid in 2+1 dimensions}
Vortices in a fluid are known to have many interesting properties. A particularly noteworthy feature is
that their position variables in two spatial dimensions (or the two transverse position variables
in three dimensions) form a canonically conjugate set, a result going back to Kirchhoff \cite{Lamb}.
Recently Wiegmann, and Wiegmann and Abanov, studied the hydrodynamic description of a large number of vortices in 2+1 dimensions, in the physical contexts of superfluids and the quantum Hall effect \cite{WA}.
The dynamics of this vortex fluid can be extracted  from the Kirchhoff description of individual vortices and the dynamics of the underlying fluid. The number density to be used for the vortex fluid is related via a constitutive-type equation to the vorticity of the underlying fluid. A background overall rotation is introduced to cancel the vorticity to a large extent so that a separation of scales, with the underlying fluid having fast dynamics and the vortex fluid as the system of slow dynamics, is possible.
Here we will not discuss more details of how the vortex fluid dynamics is extracted, for that the reader is referred to the papers cited, but we will give the key results relevant to comparison with our work.

In 2+1 dimensions, it is natural to use complex coordinates $z, \bz = x_1 \pm i x_2$, with derivatives
$\del , \bdel = {\half} (\del_1 \mp i \del_2)$.
The holomorphic component of the fluid velocity is taken as
$u = u_1 -i u_2$. The algebra of various observables can be summarized by the commutation rules
\beqar
[u (x) , \rho (y) ] &=& -i \,\del\, \delta (x-y) \nonumber\\
{} [ u (x), u^\dagger (y) ]&=& - {2\pi \over \nu}\, \delta (x-y)
\label{diff54}
\eeqar
where $\nu$ is a constant related to the strength $\Gamma$ of the individual vortices via
$\nu = 1/\Gamma$. (It may also be interpreted as the filling fraction in the context of Hall effect.)
Diffeomorphisms are generated by the operators
\beq
P = \rho \, u + {1\over 2 \nu} \del \rho , \hskip .3in
P^\dagger = u^\dagger\, \rho - {i \over 2 \nu} \bdel \rho
\label{diff55}
\eeq
Defining $P(w) = \int w \,P$, $P^\dagger (w) = \int \bw P^\dagger$, with 
complex test functions $w$, $\bw$, the commutation rule for $P$, $P^\dagger$ can be 
obtained
from (\ref{diff54}) as
\beqar
[P(w), P^\dagger (w')] &=& i \int \left( \bw' \bdel w \, P - w \del \bw' \, P^\dagger \right)
+ \int \bw' w \left( i \rho (\bdel u - \del u^\dagger ) - {2\pi \over \nu} \rho^2 \right)\nonumber\\
&&\hskip .2in + {1\over 2 \nu} \int \left(
\bw' \bdel w\, \del \rho + w \del \bw' \,\bdel \rho
+ \rho ( w\,\bdel \del \bw' + \bw' \,\bdel \del w ) \right)
\label{diff56}
\eeqar
Further simplification of the right hand side can be done using the constitutive relation
from \cite{WA}. It is given by
\beq
i (\bdel u - \del u^\dagger ) \equiv (\nabla \times u) = {2 \pi \over \nu} (\rho - {\bar \rho} )
\label{diff57}
\eeq
where ${\bar\rho} = \nu \Omega/ \pi$, with $\Omega$ being the angular velocity of the overall rotation.
The commutation rule (\ref{diff56}) now becomes
\beq
[P(w), P^\dagger (w')] = i \int \left( \bw' \bdel w \, P - w \del \bw' \, P^\dagger \right)
- {2\pi \over \nu} \int \bw' w\, \rho {\bar\rho} - {1\over \nu} \int \bdel w\, \del \bw' \, \rho
\label{diff58}
\eeq

In comparing this result with what was obtained in (\ref{diff51}), we first note that, in two 
spatial dimensions, we have the freedom of adding to $T(\xi )$ a term proportional to the
density, so that we can consider the more general
quantity
\beq
{\tilde T} (\xi ) = T(\xi ) + b \int (\nabla\times \xi) \, \rho
\label{diff59}
\eeq
where $b$ is a constant. This is essentially the same as the addition of $\del \rho$,
$\bdel \rho$ terms in defining
$P$, $P^\dagger$ as in (\ref{diff55}). The Poisson algebra for ${\tilde T}(\xi)$ can be easily worked out from (\ref{diff51}) and the other relations given in the last section as
\beq
[{\tilde T} (\xi), {\tilde T} (\xi')] =
{\tilde T} ([\xi,\xi']) - \int \left( {\rho {\bar\rho} \over \rho - {\bar\rho} } \right)\, \omega\,
\epsilon^{ij} \xi_i \xi'_j 
 - b \int \rho\,
\epsilon^{ij} \left( \del_k \xi_i \del_k \xi'_j + \del_i \xi_k \del_j \xi'_k \right) 
\label{diff60}
\eeq
where $\omega = \del_1 v_2 - \del_2 v_1$ is the two-dimensional vorticity.
We now introduce test functions $w$, $\bw$ via
$w ,\,  \bw = \xi_1 \pm i \xi_2$
so that
\beqar
{\tilde T} (\xi) &=& - \int (w\, P + \bw\, P^\dagger ) \nonumber\\
P&=& - {\half } ( {\tilde T}_{01} - i{\tilde T}_{02} ) , \hskip .2in
P^\dagger =  - {\half } ( {\tilde T}_{01} +i  {\tilde T}_{02} )
\label{diff61}
\eeqar
It is a bit tedious but straightforward to write (\ref{diff60}) in terms of the complex test functions.
We find
\beq
[P(w), P^\dagger (w')] = i \int \left( \bw' \bdel w \, P - w \del \bw' \, P^\dagger \right)
+ {1\over 2} \int \left( {\rho {\bar\rho} \over \rho - {\bar\rho} } \right)\, \omega\,
\bw' w + 2 b \int \rho\, \bdel w \del\bw'
\label{diff62}
\eeq
We have also converted our Poisson bracket relations to commutators for operators by the appropriate multiplication  by $i$, for ease of comparison. For us, the  velocity of the fluid 
obeys the commutation rule
$[ \rho v_i (x) , \rho (y)] = i \rho \del_{i} \delta (x-y)$. In comparing this with
(\ref{diff54}), we see that we must make the identification
$u \equiv u_1 -i u_2 = -\half (v_1 - i v_2 )$, which leads to
$\nabla\times u = - \half \omega$.  The constitutive relation (\ref{diff57}) in our notation is thus
\beq
\omega = - {4\pi \over \nu} \, (\rho - {\bar\rho}) 
\label{diff63}
\eeq
When this relation is used in (\ref{diff62}), we see that we have exact agreement with (\ref{diff58}),
with $b = - {1\over 2 \nu}$.

What we have shown in this subsection may be summarized as follows.
Consider the action
\beqar
S &=& -i \int \rho\, \Tr (\sigma_3 g^{-1} \del_0 g )  + i k \int \Tr (\sigma_3 g^{-1} d g ) \wedge \Omega
- \int dt \, H\nonumber\\
&&\hskip .2in + \int  A_0 dt \left[ 
i \Tr [ \sigma_3 (g^{-1} dg)^2] + {4 \pi \over \nu} (\rho - {\bar \rho}) \right]
\label{diff64}
\eeqar
where $k = 1/(\pi \Gamma) = \nu/\pi$. The two-form $\Omega$ (or its component $\Omega_{12}$)
is to be interpreted as the
angular velocity of overall rotation, and ${\bar\rho} = k\, \Omega_{12}$ as in
(\ref{diff40}). The last term in (\ref{diff63}) has a Lagrange multiplier field
$A_0$, which enforces the constitutive relation (\ref{diff63}).
Our result is that this action (\ref{diff64}) describes the effective fluid dynamics of a vortex
fluid in 2+1 dimensions; it leads to the commutation rules (\ref{diff58}) or (\ref{diff62}).
What we have obtained is thus an action formulation for the extended algebra (\ref{diff58}),
in much the same way as the topological term of the WZW model leads to the central extension
of the Kac-Moody algebra \cite{WZW}.

We close this section with a comment clarifying the comparison with \cite{WA}. 
The commutation rules (\ref{diff54}) are exactly those given in
\cite{WA}, so the algebra (\ref{diff58}) follows by direct computation.
However, the extension 
term as displayed in \cite{WA} is slightly different from the term in (\ref{diff58}),
involving the Laplace operator rather than holomorphic and antiholomorphic derivatives
on $\bw'$ and $w$. We expect that the reason for this is the following.
In \cite{WA}, the quantization of the fluid is considered
where the ground state obeys the condition $P \ket{0} = \bra{0} P^\dagger = 0$.
This is like a holomorphicity condition and,
effectively, should be equivalent to a holomorphicity condition on the test
function $w$. The
algebra given in \cite{WA}, written out with such test functions, then reduces to (\ref{diff58}).

\subsection{The vortex fluid in 3+1 dimensions}
It is also interesting to consider a vortex fluid in 3+1 dimensions
with a constitutive relation similar to (\ref{diff63}).
The second extension to the algebra (\ref{diff58}) arising from the addition of
$\int \rho (\nabla \times \xi )$ to $T(\xi)$ is irrelevant for this case; it is trivial from the point of view of the cohomology of the algebra anyway, so we will focus on (\ref{diff51}). We write the vorticity as
\beq
\omega_{ij} = \del_i v_j - \del_j v_i = \epsilon_{ijk} N^k \, \omega
\label{diff65}
\eeq
where $N^k$ is a unit vector giving the orientation of the vorticity at a given point
and $\omega$ is its magnitude. Unlike in two dimensions, we now have vortex lines, so
$N^k$ gives the local orientation of a set of vortex lines coarse-grained over a small volume.
 As in the (2+1)-dimensional case, we
expect $\omega$ to be proportional to the number density of vortices. So we propose to use the
same constitutive relation in three dimensions as well, namely,
\beq
\omega = - {4\pi \over \nu} \, (\rho - {\bar\rho}) 
\label{diff66}
\eeq
where ${\bar \rho}$ is given in terms of the 3-form $\Omega$ as in (\ref{diff40}).
The algebra  (\ref{diff51}) takes the form
\beqar
[ T(\xi), T(\xi')] &=& T ([\xi, \xi']) + \int  \epsilon_{ijk}\, \xi^i \xi'^j c^k\nonumber\\
c^k &=&  {4 \pi \over \nu}  \rho {\bar\rho} \, N^k
\label{diff67}
\eeqar

We will now relate this to some recent work on the algebra of vector fields for 
an incompressible fluid \cite{Raj}.
If we consider the reduction of the algebra (\ref{diff67}) to the incompressible case,
with the vector fields $\xi^i$, $\xi'^j$ being divergence-free,
$\rho \, {\bar \rho}$ can be taken to be a constant.
The motion of the vortices is on a two-dimensional surface
transverse to their vortex lines, i.e., transverse to the vector $N^k$.
If we have a large dense collection of vortices, $N^k$ will be uniform in the
transverse surface, just as it was in 2+1 dimensions. 
As one follows along the vortex lines, $N^k$ can change orientation.
It is useful to consider the case of $N^k$ being constant, independent of $\vx$.
This would be realizable at least in some subvolume of space.
In this case, the extension term in (\ref{diff67}) becomes
\beq
{\rm Extension} = c^k \int \epsilon_{ijk} \xi^i \xi'^j
\label{diff68}
\eeq
If space is taken to be a 3-torus as in \cite{Raj}, one can parametrize the divergence-free
vector fields as
\beq
\xi^i  = \epsilon^{iab} \alpha_a m_b \, e^{i {\vec m} \cdot \vx} ,
\hskip .2in
\xi'^j  = \epsilon^{j rs} \beta_r n_s \, e^{i {\vec n} \cdot \vx}
\label{diff69}
\eeq
Here $m_i$, $n_i$ are vectors of integer components and
$e^{i {\vec m}\cdot \vx}$, $e^{i {\vec n}\cdot \vx}$ provide a basis for functions on the torus.
By taking the components of each of
$\alpha_i$, $\beta_i$ and $({\vec\alpha} \times {\vec\beta})_i$ to be linearly
independent over the integers, one can get a dense set of test functions.
With the test functions in (\ref{diff69}), the extension term (\ref{diff68}) becomes
\beq
{\rm Extension} = - ({\vec\alpha} \times {\vec\beta})\cdot {\vec n} ~
{\vec c} \cdot {\vec n}
\label{diff70}
\eeq
This is agreement with the central extension considered in \cite{Raj}. 
So our conclusion is that the topological term $I_3$ can explain the
central extension of \cite{Raj} as a special case
 with the reduction conditions as explained above.
The consideration of a constant $c^k$ was just for showing
this connection.
But using the action (\ref{diff39})
we can go beyond considering constant $c^k$, with the more general
algebra (\ref{diff67}) being applicable to a vortex fluid in 3+1 dimensions.
The extension is no longer a central term in the general case.
\section{The term $I_2$ in 3+1 dimensions}
We now turn to the second case of a topological term
we listed at the end of section 3, namely, $I_2 = \Tr (g^{-1} dg)^3 \wedge C$ where $C$ will be taken as external, with $d C \neq 0$. Adding such a term with coefficient
$-k/3$, the action we are considering is
\beq
S = - i \int \rho \, \Tr ( \sigma_3 g^{-1}\del_0 g ) - {k \over 3} \int 
\Tr (g^{-1} dg )^3 \wedge C - \int dt \, H
\label{diff71}
\eeq
The terms with the time-derivative of $g$ lead to the canonical 1-form and 2-form
\beqar 
\A&=& - i \int \rho\, \Tr (\sigma_3 g^{-1}\delta g ) - k \int \Tr [ g^{-1}\delta g\, (g^{-1} dg)^2] \wedge C
\nonumber\\
\delta \A &=& -i \int \left[  \delta \rho\, \Tr (\sigma_3 g^{-1}\delta g) - \rho \Tr (\sigma_3 (g^{-1} \delta g )^2 \right] + k \int \Tr [ (g^{-1}\delta g)^2 g^{-1} dg ] \wedge C
\label{diff72}
\eeqar
In this case the identification of the Hamiltonian vector fields is not very easy, it is simpler to work out
the inverse of the canonical 2-form and form the Poisson brackets.
Towards this, we write
\beqar
g^{-1} \delta g &=& -i t_a \E^a = -i t_a \E^a_\alpha \delta \vf^\alpha \nonumber\\
g^{-1} dg &\equiv& -i t_a I^a
\label{diff73}
\eeqar
The canonical 2-form $\delta \A$ is then given as
\beq
\delta\A = - \delta \rho\, \E^3 + {1\over 2} \rho \, \epsilon_{ab3} \E^a \wedge \E^b
- {1\over 2} \epsilon_{abc} b^c \, \E^a \wedge \E^b
\label{diff74}
\eeq
where 
\beq
b^c = {k \over 2} B_i I^c_i , \hskip .3in
B^k = \epsilon^{klm} \del_l C_m
\label{diff75}
\eeq
Taking the inverse of $\delta\A$, the Poisson bracket of functions $A$, $B$ is given as
\beqar
[A, B]&=& \int \biggl[{\delta A \over \delta \rho(x)} \E^{-1\alpha}_3 {\delta B \over \delta \vf^\alpha(x)}
- {b_1 \over \rho - b_3} {\delta A \over \delta \rho(x)} \E^{-1\alpha}_1 {\delta B \over \delta \vf^\alpha(x)}
- {b_2 \over \rho - b_3} {\delta A \over \delta \rho(x)} \E^{-1\alpha}_2 {\delta B \over \delta \vf^\alpha(x)}
\nonumber\\
&&\hskip .25in
- {1 \over \rho - b_3} {\delta A \over \delta \vf^\alpha} \E^{-1\alpha}_1 
\E^{-1\beta}_2{\delta B \over \delta \vf^\beta}
- (A \leftrightarrow B ) \biggr]
\label{diff76}
\eeqar
The computation of the Poisson brackets of various quantities of interest using this formula is tedious but straightforward. The bracket of $\rho$ and $g$ is given by
\beq
[ \rho (f) , g(x)] = -i \left[ {  \rho \over  \rho - b_3} \, g\, t_3 - {1 \over \rho - b_3} \,
g \, b^a t_a \right]\, f(x)
\label{diff77}
\eeq
As before, using the energy-momentum tensor obtained by varying (\ref{diff71}) with respect to the
metric, we have
\beq
T(\xi) = - i \int \rho\, \Tr (\sigma_3  g^{-1} \xi\cdot \del g) = \int \rho\, \xi\cdot v
\label{diff78}
\eeq
For this we have the Poisson bracket algebra
\beqar
[ \rho(f), T(\xi) ] &=& - \int \rho\, \xi\cdot \nabla f - \int {\rho \over \rho - b_3} \Tr \bigl([\sigma_3 , b^a t_a]
(g^{-1} \xi\cdot \del g) \bigr)\, f\nonumber\\
{}[T(\xi), T(\xi')]&=& T([\xi, \xi']) + k \int {\rho \over \rho - b_3}
\Tr \bigl( [ g^{-1} \xi\cdot \del g, g^{-1} \xi'\cdot \del g] g^{-1} dg\bigr) \wedge d C
\label{diff79}
\eeqar
Recall that the helicity of the  fluid is given by
\beq
\C = {1\over 8\pi} \int v \cdot \omega = {1\over 12 \pi} \int \Tr (g^{-1} dg)^3
\equiv \int \sigma
\label{diff80}
\eeq
In terms of the density $\sigma$ for helicity as defined above, the bracket relation for
$T(\xi)$ can be written as
\beq
[T(\xi), T(\xi')] =  T([\xi, \xi'])  + 4\pi k \int \left( {\rho\, \sigma \over \rho - b_3}\right) \,
({\vec \xi} \times {\vec \xi'} ) \cdot {\vec B}
\label{diff81}
\eeq

A natural question at this point would be whether there is a physical system for which the present case of $I_2$ is applicable. The example of vortex fluids discussed in the last section can be a guide in this direction. As known for a long time, helicity is related to knots and links for vortex lines.
For example, consider the linking of a vortex line with another, the latter forming a circle which may be viewed as the boundary of some two-surface $\Sigma$. If vortices are approximated by thin lines, the integrand $\sigma$ for helicity has support at the point of intersection of the vortex line with the surface $\Sigma$. The support for $\sigma$ is point-like localized in the thin vortex approximation.
As time evolves, these points can move and for a fluid with a dense
collection of such knots one can envisage constructing an effective hydrodynamics of knots or links.
The action with the $I_2$-term added as in (\ref{diff71}) is a good candidate for
such an effective hydrodynamics.
Again an overall rotation (interpreted as $dC$) may be needed to ensure a proper separation of scales.
Although well-motivated, admittedly, this connection is still
speculative; it will need more work to tie down the specifics.
\section{Discussion}
In this paper, we considered some topological terms which can be added to the standard actions for sigma models and for fluid dynamics. The example of the sigma model with $\mathbb{CP}^2$ as the target space shows how the additional term can lead to a conflict between diffeomorphisms for the target and base spaces.
For the case of fluid dynamics, it is worth emphasizing that we are not introducing any additional variables or degrees of freedom. We use the standard action in terms of the Clebsch variables with the topological terms added. In 2+1 dimensions, we showed how one such topological term leads to the effective hydrodynamics of a vortex fluid as derived in \cite{WA}. This provides an action-based derivation of the extended diffeomorphism algebra in much the same way as the WZW model gives an action-based derivation of the central extension of the Kac-Moody algebra. A similar analysis
was made in 3+1 dimensions, presumably applicable to a fluid of vortex lines.
A special case leads to the central extension found in \cite{Raj}.
We also discussed another topological term using the helicity of the fluid, which might apply to a fluid made of knots and links of vortex lines.

There are a couple of other relevant observations. For the example of the invariant $I_2$, we considered the one-form $C$ to be external, with $d C \neq 0$. As already mentioned, one could also consider $C$ to be of the form $\Tr (\sigma_3 h^{-1} dh)$ where $h$ defines the Clebsch variables for
another fluid. The algebra of observables for the second fluid will be similar to what was obtained for the $I_3$-invariant. However, for the combined system, there can be cross terms. These need further analysis.
Our second observation is about the Hopf invariant. If one considers vortices
of quantized charge or strength, the Clebsch variables are described by $SU(2)$ rather than
$SU(1,1)$ with the vorticity as the pull-back to space of the volume form
on $S^2 = SU(2)/U(1)$.
 It is then natural to consider the Hopf invariant in 2+1 dimensions, and in 3+1 dimensions with an additional one-form $C$. A partial analysis of the 2+1 case is given in \cite{Nair-Ray}, but more needs to be done for this case as well.

 \bigskip

I thank A. Abanov, G. Monteiro and  especially P. Wiegmann for many useful
discussions.
This research was supported in part by the U.S.\ National Science
Foundation grant PHY-1820721
and by PSC-CUNY awards.
 

\end{document}